# Rectification effect in organic junctions based on aryl thin films covalently attached to a multilayer graphene electrode

*Clément Barraud*[1], Matthieu Lemaitre[1], Roméo Bonnet[1], Jacko Rastikian[1], Chloé Salhani[1], Stéphanie Lau[2], Quyen Van Nguyen[2], Philippe Decorse[2], Jean-Christophe Lacroix[2], Maria Luisa Della Rocca[1] Philippe Lafarge[1] and Pascal Martin*[2]*

1. MPQ UMR 7162, Université Paris Diderot, Sorbonne Paris Cité, CNRS, F-75013 Paris, France

2. ITODYS UMR 7086, Université Paris Diderot, Sorbonne Paris Cité, CNRS, F-75013 Paris, France

**ABSTRACT**

The quantum interaction between molecules and electrode's material at molecules/electrode interfaces is a major ingredient in the electronic transport properties of organic junctions. Driven by the coupling strength between the two materials, it results mainly in a broadening and an energy shift of the interacting molecular orbitals. Using new electrodes materials, such as the recent semi-conducting two-dimensional nanomaterials, has become a recent challenge in the field of molecular/organic electronics that opens new possibilities for controlling the interfacial electronic properties and thus the charge injection properties. In this article, we report the use of atomically thin two-dimensional multilayer graphene films as base electrode in organic junctions with a vertical architecture. The interfacial electronic structure dominated by the covalent bonding between bis-thienyl benzene diazonium-based molecules and the multilayer graphene electrode has been probed by ultraviolet photoelectron spectroscopy and the results compared with those obtained on junctions with standard Au electrodes. Room temperature injection properties of such interfaces have been also explored by electronic transport measurements. We find that, despite strong variations of the density of states, the Fermi energy and the injection barriers, both organic junctions with Au base electrodes and multilayer graphene base electrodes show similar electronic responses. We explain this observation by the strong orbital coupling occurring at the bottom electrode/ bis-thienyl benzene molecules interface and by the pinning of the hybridized molecular orbitals.

**INTRODUCTION**

The architecture of the two- and three-terminals building blocks of organic electronic devices, such as electroluminescent diodes and organic transistors, is largely inspired by their inorganic counterparts. Practically, the control and understanding of the electrode/molecules interface's properties is crucial for reaching high performances and designing functionalities[1], such as electrical rectification. Symmetric current density as a function of bias voltage ($J$-$V$) characteristics were widely observed in two-terminal devices and discussed in the case of single molecule junctions[2], monolayers[3] or oligomers based-organic junctions, as in the case of a large variety of aromatic nanostructures between conducting carbon electrodes[4,5]. It has been shown that to introduce asymmetry (rectification) in the electronic response of molecular and organic junctions, it is necessary to introduce an asymmetry along the transport direction[6,7]. This can be achieve either by using different contact electrodes (i.e. with different injection properties), and/or by using asymmetric anchoring moieties or molecular structures as originally proposed by Aviram and Ratner[8]. Using such artificial systems to create nano-rectifiers has been widely reported and recently reviewed[9]. The electrode/molecules coupling plays also a crucial role in defining the amplitude of the rectification ratio (RR)[9]. Molecules can be either attached to electrodes via strong (chemisorption) or weak interactions (physisorption). Depending on the nature of the interactions on both sides of the junctions, the RR has been demonstrated to vary from few tenth to few thousands[9]. Recently record RR up to $10^5$ were reported in alkyl-ferrocene-based self-assembled monolayers contacted by chemisorption to a bottom electrode and physisorption to a top electrode[10]. The rectification effect was also shown to be strongly dependent upon the position of the Fc unit (i.e. the molecular structure) between the two electrodes[11]. We have shown before that based on diazonium approach oligo-BTB form high quality organic layer and lead to highly stable organic junctions. We have reported before molecular diodes based on diazonium grafting in junctions of the form of Au-BTB//Ti/Au where "-" and "//" denote the bottom interface and the top interface respectively. Those organic diodes presented large rectification ratio of 1000 and we have systematically studied the mechanism of charge transport across this layer.

In this article, we focus on the electrode's properties. New atomically thin electrodes made of two-dimensional materials, such as graphene, have been recently proposed and developed to replace more traditional metallic electrodes[12] more sensitive to oxidation and/or electro-migration. Graphene, multilayer graphene (MG) and graphite were recently used as a soft, transparent and tunable electrode for contacting rubrene single crystals[13], self-assembled monolayers[14–16] and even single molecules[17–22]. The strength of the chemical bonding between a graphene/graphite electrode and molecules was demonstrated to be a key ingredient over the global electronic response of single-molecule junctions[21].

Here we demonstrate electrical rectification effects in the case of MG-(oligo(1-(2-bisthienyl) benzene, named BTB) diazonium-based thin film//Ti/Au junctions. We show that rectification is directly linked to the strong orbitals coupling at the MG-BTB interface like in the well-studied case of Au-BTB interfaces[23] and to pinning of the hybridized molecular orbitals at top and bottom electrode/molecules interfaces. Our observation is strongly supported by a comparative analysis of those two device's interfaces by electronic transport measurements and ultraviolet photoelectron spectroscopy.

## RESULTS AND DISCUSSION

MG-BTB//Ti/Au junctions were fabricated by first transferring a MG layer (purchased from Graphene Supermarket®) over an insulating 280 nm $SiO_2$ substrate. The transfer technique is adapted from Ref. 24 and is described in the experimental section. Two Raman spectra acquired on MG after transfer (laser wavelength: 633 nm) in two different locations are shown in figure 1(a), together with an optical microscope image (inset of figure 1(a)) of a 500 nm $SiO_2$ substrate covered by the transferred MG. The observed changes in the shape and relative heights and widths of the G and 2D peaks reveal that the number of graphene layers is not uniform over the entire MG film[25]. This is expected as the MG thickness is specified from Graphene Supermarket® to be equal to 4 monolayers in average with local variations between 1 and 7 monolayers. The absence (black line) or the weak (red line) D peak arising around 1350 $cm^{-1}$ confirms the excellent structural quality of the MG after transfer. The MG is then patterned by optical lithography and $O_2$ plasma (70 W during 120 s) in long stripes of 10 mm long and 40 µm wide.

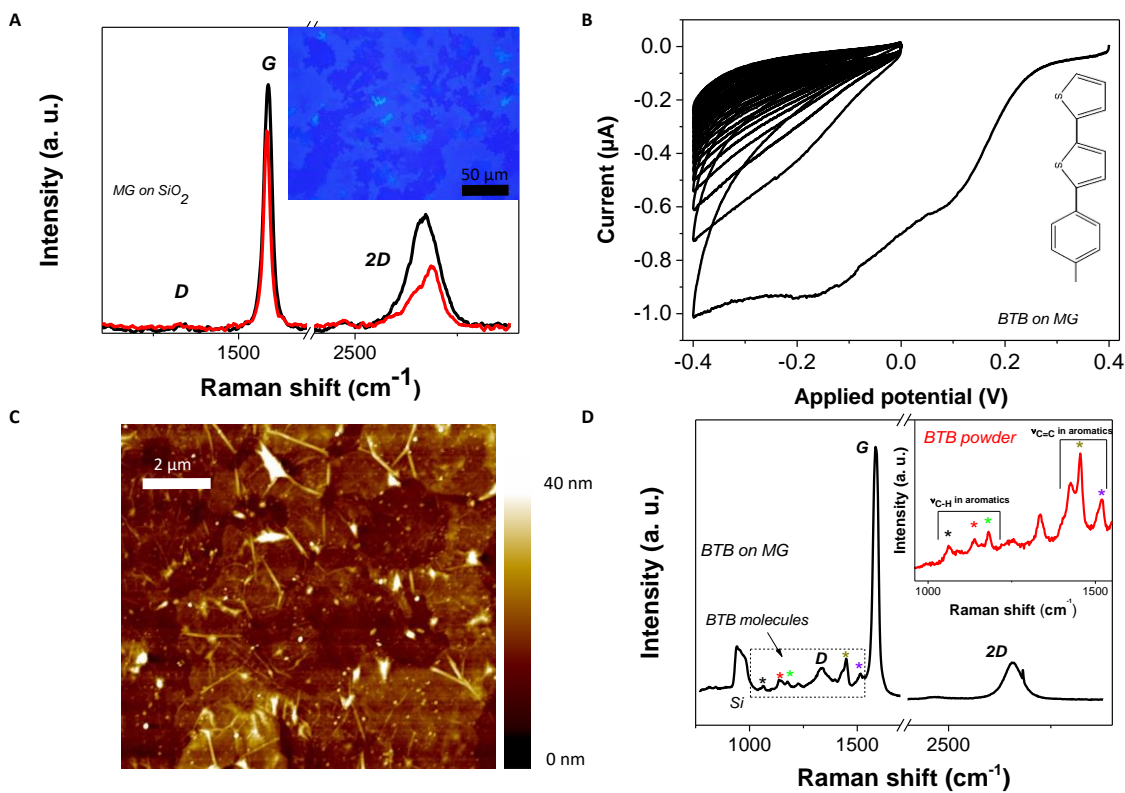

Figure 1. (a) Two Raman spectra acquired at different locations on MG after transfer on a 280 nm $SiO_2$ substrate (laser wavelength: 633 nm). Inset: optical image of a MG film deposited over a 500 nm $SiO_2$ substrate. (b) Cyclic voltammograms of (1-(2-bisthienyl)-4-aminobenzene) (BTAB) electro-reduction in acetonitrile (10 cycles at 0.1 $V.s^{-1}$) over the MG electrode. Inset: molecular structure of the BTB molecule. (c) Atomic force microscopy topography image of a MG grafted with BTB molecules acquired in ambient conditions in tapping mode (10 x 10 $\mu m^2$) (d) Raman spectrum acquired on MG/BTB surfaces highlighting the region between 1000 $cm^{-1}$ and 1550 $cm^{-1}$ where the resonances associated to the oligo(BTB) molecules are observed. Inset: Raman spectrum acquired on a high-purity 1-(2-bisthienyl) benzene amine powder revealing similar resonances.

After fabrication of the patterned MG electrodes, electrochemical grafting of BTB was carried out in solution by cyclic voltammetry. Figure 1(b) shows a cyclic voltammetry characteristic of the reduction of the 1-(2-bisthienyl) benzene diazonium cation at the surface of a MG electrode. It shows an irreversible wave during the first scan [+0.4 V; -0.4 V], which corresponds to the formation of aryl radicals[26] at the proximity of the MG electrode. In the following scans, the current strongly drops to zero as the conducting surface of the electrode becomes passivated by the growth of an oligo-BTB thin layer[27]. BTB thin films are electro-active and can be easily p-doped at a potential close to 0.5 V/SCE (saturated calomel electrode)[27]. As a consequence, the conductance of a BTB thin film can be switched and diode-like behavior with a high rectification ratio is observed in the electrochemical response of outer-sphere redox probes[28]. At the end of the process, the MG surface is fully covered by the organic film as previously detailed for graphene and graphite surfaces[29]. No trace of diazonium salt was detected by X-ray photoelectron spectrometry analysis within the resolution of the detector (not shown). The final thickness and roughness of the BTB film were determined by atomic force microscopy as shown in figure 1(c) and was found to be around 10 ± 1 nm. After electro-grafting, the MG electrode was again characterized by Raman spectroscopy. Figure 1(d) shows the Raman spectrum (laser wavelength: 633 nm) acquired on the MG/BTB surface. While the main features from the graphitic structure, the D, G and 2D resonances, can still be identified, additional resonances (individually marked with colored *) are observed in the frequency range 1000 $cm^{-1}$ and 1550 $cm^{-1}$. They are attributed to vibrational modes (stretching vibration for C=C aromatics (1520 $cm^{-1}$ and 1450 $cm^{-1}$) and C-H aromatics (1060 $cm^{-1}$, 1140 $cm^{-1}$ and 1180 $cm^{-1}$)) within the oligo-BTB film[30,31]. This is further probed by measuring the Raman spectrum of high-purity 1-(2-bisthienyl) benzene amine powder (inset of figure 1(d)), where clearly resonances at identical frequencies are observed. In the Raman spectra of figure 1(d), the $sp_2$ graphitic structure are still evident as only the topmost layer of MG is chemically functionalized[29] leaving the remaining layer of the MG structure intact. The topmost layer is $sp_3$ hybridized giving rise to the increase of the visibility of the D peak around 1350 cm$^{-1}$. This result is confirmed by measuring the resistance (not shown) of a 2.5 mm long - 40 µm wide functionalized MG stripe. A value of ~25 kΩ is extracted in good agreements with standard electrical characterizations of MG layers[32]. The top electrode is finally patterned in cross-bar geometry by optical lithography. A thin film of Ti (1 nm)/Au (50 nm) is then evaporated at pressure < 10$^{-7}$ mBar and with low rate < 0.1 nm/s leading to the MG-BTB//Ti/Au organic/inorganic heterostructures without titanium oxide formation.

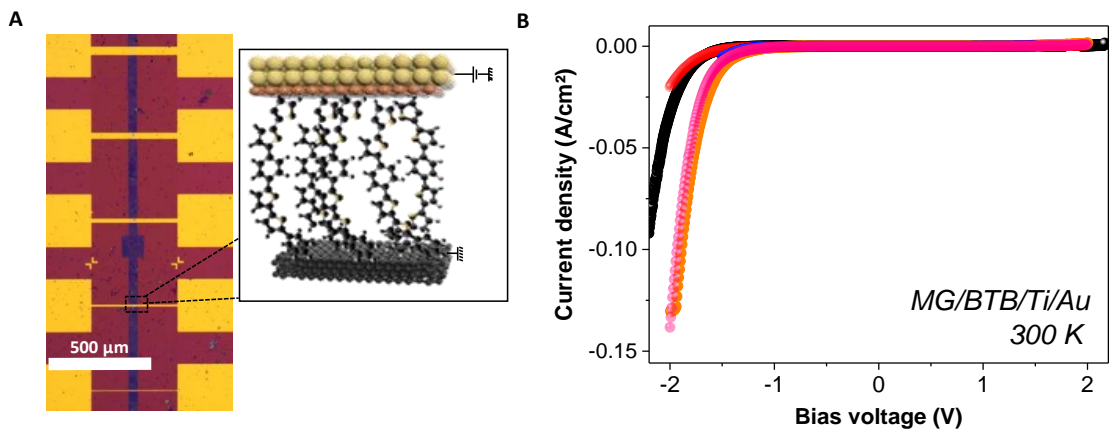

Figure 2. (a) Optical image of MG-BTB//Ti/Au junctions with various dimensions of junctions. A schematic view of a polarized junction is presented in the black box. (b) Room temperature current density as a function of bias voltage characteristics acquired for five different MG-BTB//Ti/Au junctions. Dimensions of the measured junctions were ranging from 5 x 40 µm$^2$ to 20 x 40 µm$^2$.

An optical image of the devices accompanied by a schematic of a polarized junction is depicted in figure 2(a). All the junctions were electrically characterized using standard DC measurement techniques. Details about the experimental setup can be found in the experimental section. Room temperature *J-V* characteristics for five different MG-BTB//Ti/Au junctions with area ranging from 5 x 40 µm$^2$ to 20 x 40 µm$^2$ are plotted in figure 2(b). A rectifying effect is revealed with a rectification ratio RR = |$J$(-2 V)/$J$(+2 V)| up to 100. The current density in the conducting states at negative voltages (i.e. current flowing from the bottom electrode to the top electrode) reaches values up to 0.15 A/cm$^2$.

To gain insights on the electronic properties of the different hybrid interfaces, we have performed ultraviolet photoelectron spectroscopy (UPS) of a bare MG surface and of a MG-BTB interface. As mentioned already, BTB molecules are known to be *p*-type molecules and transport occurs preferentially through the occupied molecular orbitals[5,23]. For UPS spectroscopy the MG-BTB bilayer was fabricated following the same procedure with a thinner BTB thickness (3-4 nm) in order to get relevant information for the injection barrier. The secondary electron cut-off and valence band spectra for MG (black line) and for MG-BTB (red line) are represented in figures 3(a) and 3(b) respectively. The MG spectra is in good agreement with the literature[11]. The very small density of states at the Fermi energy of MG gives the observed rather weak increase of the signal towards higher binding energies compared to standard metals[33]. From these measurements, we can extract the vacuum level shift ($\Delta E = 0.2$ eV), generally interpreted as the presence of dipoles at the electrode/molecules interfaces[34] and the injection barrier 3-4 nm away from the interface ($\varphi_{MG} = 1.05$ eV). Both values of vacuum level shift and injection barrier are comparable with reported values of BTB and other diazonium compounds grafted on pyrolized photoresist film by Sayed *et al.*[4]. These values are reported on the energy diagram of the MG-BTB interface presented in figure 3(c). The first molecular monolayer is expected to be strongly coupled to MG and molecular orbitals can be broadened by several eV[35] and on resonance with the Fermi energy of the MG electrode[36]. In the specific case of diazonium molecules, a strong C-C chemical bond is expected between the molecule and the carbon-based surface[4]. In case of chemisorption of molecules on metals, orbital's broadening up to few eV have been calculated for instance for Co/Alq$_3$ surfaces[36]. It then tends to decay within the molecular layer as also shown for a Co/Alq₃ interface[36] and experimentally measured for N,N'-bis-(1-naphthyl)-N,N'-diphenyl1-1,1-biphenyl1-4,4'-diamine (α-NPD) deposited on Au[39].

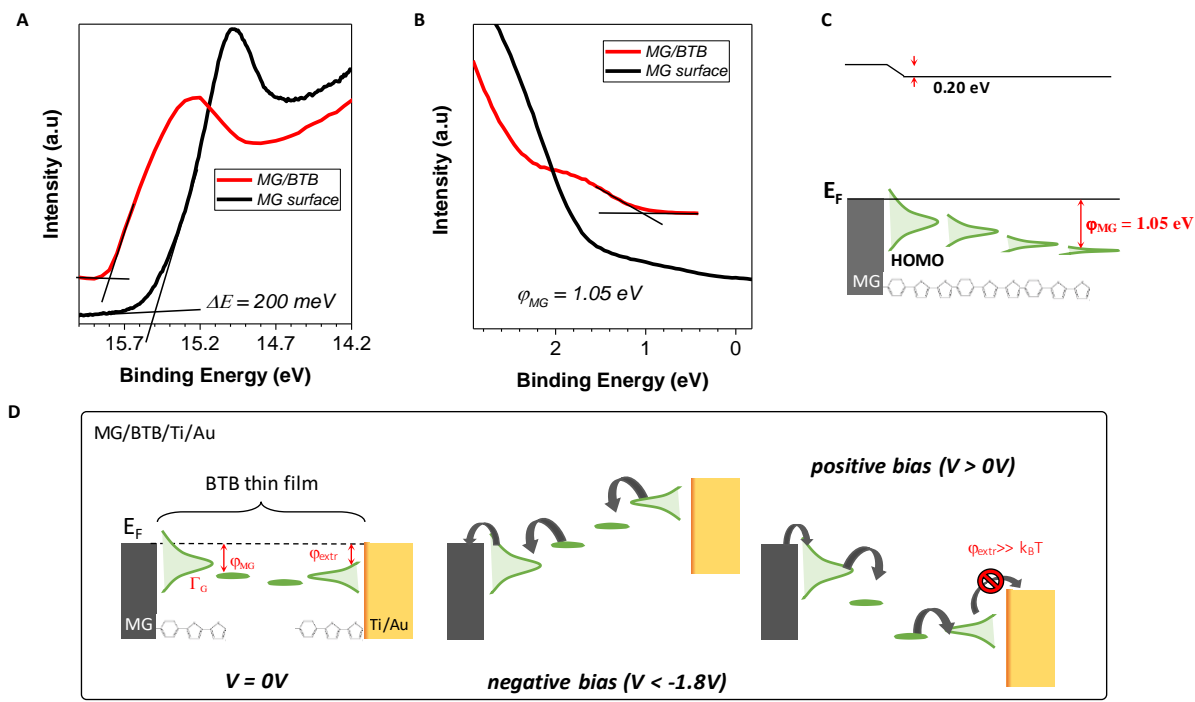

Figure 3. (a) and (b) UPS spectra acquired on a bare MG surface (black line) and on a MG/BTB surface (red line). The spectra were taken with sample biased at -5 V to clear the detector work function. (c) Schematic of the interfacial electronic structure with the reported values of dipoles and barrier. The HOMO of BTB is strongly broadened and shifted in contact with the MG electrode. The orbital broadening decays with respect to the distance from the electrode. (d) Transport mechanism for a MG-BTB//Ti/Au junction. (left) Electronic structure of the junction at zero bias voltage. (middle) At negative bias voltages in the passing state. (right) At positive bias voltages. In this last case, the extraction energy barrier (φ_extr) is higher than the thermal energy preventing the current to flow.

The top BTB//Ti/Au interface has been recently characterized by XPS highlighting the presence of carbide moieties[37] and a HOMO broadening is thus expected. Based on electronic transport measurements and UPS analysis, we propose in figure 3(d) a transport mechanism for MG-based organic junctions at zero, negative, and positive bias voltage that explains the rectification observed in MG-based organic junctions. Pinning of the HOMO at both interfaces is considered with a strong broadening at the MG interface, this results in a current flow at negative bias voltage below -1.8 V. On the opposite, at the BTB/Ti/Au top interface, the extraction energy barrier is higher than the thermal energy preventing the current to flow at positive bias voltage[37]. Hopping transport via polaronic states is considered in between the represented interfacial molecular orbitals[23]. Note that the LUMO level is not represented due to the large HOMO-LUMO gap (3.1 eV) of the BTB molecule[37].

To highlight the effect of the orbital coupling, we compare now the *J-V* characteristics presented in figure 2(c) to *J-V* characteristics measured in the same conditions on Au-BTB//Ti/Au, with BTB film of identical thicknesses. We show in figure 4(a) and 4(b) a schematic of an Au-BTB//Ti/Au junction and four different room temperature *J-V* characteristics. They clearly show rectification effect similar to that of MG based electrode junctions in figure 2(c), with RR up to 170* at ± 2 V and current densities up to 0.3 A/cm². The similar electronic responses of the two systems suggests that the coupling intensity at the MG-BTB interface is similar to the one observed at Au-BTB interfaces, opening for diazonium-based organic films over MG electrodes as a possible platform for molecular junctions. We have thus performed UPS analysis of Au surfaces and Au-BTB interfaces to compare with the data extracted from MG-based surfaces. The secondary electron cut-off and valence band spectra are given in figures 4(c) and (d). The Au surface spectra (black line) is in agreement with literature[33,38]. UPS spectra show very sharp steps at the Fermi energy due to the strong density of states present in the metal in opposition to MG (figure 3(b) – black line). This sharp step disappears as soon as the BTB film is inserted due to the much lower density of states of the organic layer. The measured injection barrier of Au-BTB interface is ∼ 1.2 eV, approximately 150 meV higher than the one measured for the MG/BTB case. The values of interfacial dipoles is ∼ 1.6 eV so largely different compared to the MG-BTB case but comparable with pentacene, Alq₃ or α-NPD[39] or thiols on Au electrodes[40]. These values are reported on the schematics in figure 4(e).

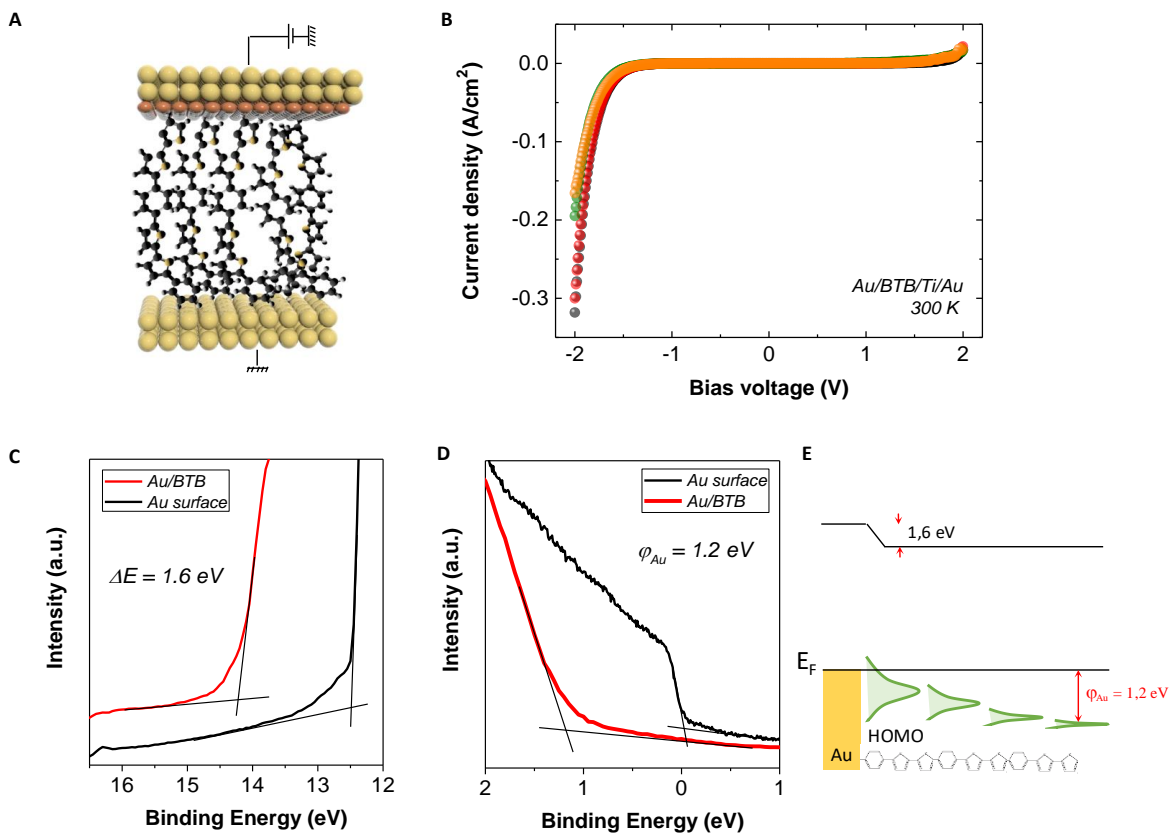

Figure 4. (a) Schematic of an Au-BTB//Ti/Au organic junction. (b) Room temperature current density as a function of bias voltage characteristics acquired for four different Au-BTB//Ti/Au junctions. Dimensions of the junctions are 20 x 20 μm². (c) and (d) UPS spectra acquired on a bare Au surface (black line) and on an Au-BTB surface (red line). The spectra

were taken with sample biased at -5 V to clear the detector work function. (e) Schematic of the interfacial electronic structure with the reported values of dipoles and barrier.

This comparative study between the MG-BTB//Ti/Au and the Au-BTB//Ti/Au devices reveals surprisingly that the electronic responses (*J-V*) are almost independent of the bottom electrode nature, of its density of states at the Fermi energy, of its interfacial dipole with the molecules and of the injection barriers. It thus partially rules out the recent mechanism proposed to explain rectification reported in a slightly different system: graphite/amino-benzene-based molecule/Au tip junction under a scanning tunneling microscope[21]. This mechanism relies on the rapidly varying and highly dispersive nature of the density of states of graphite around the Fermi energy. Focusing on the electrode/molecule interface, our results agree more with the recent work of Sayed *et al.* in similar systems. They have shown that the injection barrier of PPF/diazonium-based molecules interfaces was independent of the molecular nature of the grafted layer[4]. The effect was attributed to a dubbed "compression" of the molecular orbitals. Our results go a step further in this interpretation by demonstrating that the strong electronic coupling given by the BTB-diazonium moieties dominates the injection properties of the base electrode/BTB interface.

**CONCLUSIONS**

We demonstrated the use of atomically thin multilayer graphene as a base electrode for molecular junctions. A clear rectification effect is reported in MG-BTB//Ti/Au junctions at room temperature comparable to what observed in Au-BTB//Ti/Au junctions. This is explained by the strong orbital coupling present at the MG-BTB interface and by the pinning of hybridized molecular orbitals at the MG-BTB and BTB//Ti/Au interfaces. A more detailed BTB-thickness dependent UPS study should be needed in order to get the evolution of the position and broadening of the different unoccupied molecular orbitals. This work is a further step towards all carbon molecular junctions which are already proved to more stable than those using classical Au electrodes[4.]

EXPERIMENTAL SECTION

**Multilayer graphene transfer**. MG was purchased on Ni films. The Ni/MG bilayer was first covered with a thin layer of PMMA 950K deposited by spin-coating (4000 rpm) and baked at 110°C for 10 minutes. The Ni/MG/PMMA trilayer was then gently deposited at the surface of a $FeCl_3$ bath in order to dissolve the Ni substrate. After dissolution, the floating MG/PMMA bilayer was transferred in a deionized $H_2O$ bath for 5 minutes. MG was cleaned in successive baths following the recipe described in reference 24 (i.e. RCA clean). It was then transferred in a deionized $H_2O$ bath and then gently "fished" by the substrate. After transfer, the sample was slowly baked from room temperature to 110°C in order to evaporate the remaining water. The PMMA film was finally removed by immersing the sample in a warm (50°C) acetone bath for 5 minutes and in an isopropanol bath for 2 minutes.

**Electro-grafting**: The 1-(2-bisthienyl)-4-aminobenzene (BTB) was synthesized following a published procedure[27]. For electrochemical experiments, a conventional one-compartment, three-electrode cell was employed. A CHi 760C potentiostat (CH Instruments, Austin, TX) was used. The auxiliary electrode was a platinum grid. A SCE (3 M KCl) in a ACN/LiClO$_4$ bridge was used as a reference electrode.

**Experimental setup**. DC electrical measurements were performed by applying a DC voltage signal to the junction top electrode by a low-noise voltage source (Yokogawa GS200). The output current was measured by a low-noise current-voltage amplifier connected to the bottom electrode. The amplified output was measured by a digital voltmeter (Agilent 34405A).

**Spectroscopy characterizations:** XPS and UPS analyses were performed in ultrahigh vacuum system (VG Scientific ESCALAB 250) with a base pressure of $2.10^{-10}$ mbar. XPS was performed using an Al KR X-ray source (1486.6 eV) and a microfocused monochromatic and magnetic lens. The spectra were acquired in the constant analyzer energy mode with pass energies of 150 and 40 eV for the survey and the narrow regions, respectively. UPS was carried out using monochromatic He I (21.21 eV) emission together with a toroidal mirror monochromator.

**Yield and reproducibility**: A high percentage (80 %) of working and reproducible devices is found in the case of Au-BTB//Ti/Au junctions in agreement with previous reports[37]. The yield of working devices is reduced when MG is inserted (around 20 %). It may be explained by the relatively high peak-to-peak roughness (< 10 nm) due to folds of the MG electrode compared to a more standard evaporated Au electrode (< 1 nm) as revealed by the AFM image presented in the figure 1(c). All the *J-V* characteristics presented are stable over periods of several hours.

**Corresponding Authors**

clement.barraud@univ-paris-diderot.fr and pascal.martin@univ-paris-diderot.fr

**Contributions**

 All authors have given approval to the final version of the manuscript.

**ACKNOWLEDGEMENTS**

We acknowledge C. Manquest, P. Filloux and S. Suffit for technical supports within the clean-room of the Laboratoire Matériaux et Phénomènes Quantiques (UMR 7162) at the Université Paris Diderot and Dr M. Marsi at the université Paris Sud for helpful discussions about UPS data. This work is supported by the 2DSPIN project from the "Ville de Paris" Emergence program. ANR (Agence Nationale de la Recherche) and CGI (Commissariat à l'Investissement d'Avenir) are gratefully acknowledged for their financial support of this work through Labex SEAM (Science and Engineering for Advanced Materials and devices) ANR 11 LABX 086, ANR 11 IDEX 05 02. This work has been also supported by the Region Ile-de-France in the framework of DIM Nano-K (SMS project).

ABBREVIATIONS

MG, multilayer graphene; BTB, (1-(2-bisthienyl)benzene); AFM, Atomic Force Microscopy.

REFERENCES

(1)   Cahen, D.; Kahn, A.; Umbach, E. Energetics of Molecular Interfaces. *Mater. Today* **2005**, *8*, 32–41.

(2)   Taniguchi, M.; Tsutsui, M.; Mogi, R.; Sugawara, T.; Tsuji, Y.; Yoshizawa, K.; Kawai, T. Dependence of Single-Molecule Conductance on Molecule Junction Symmetry. *J. Am. Chem. Soc.* **2011**, *133*, 11426–11429.

(3)    Shi, X.; Dai, Z.; Zeng, Z. Electron Transport in Self-Assembled Monolayers of Thiolalkane: Symmetric    $I-V$    Curves and Fano Resonance. *Phys. Rev. B* **2007**, *76*, 235412.

(4)    Sayed, S. Y.; Fereiro, J. A.; Yan, H.; McCreery, R. L.; Johan, A.; Bergren, A. J.; Johan, A. Charge Transport in Molecular Electronic Junctions: Compression of the Molecular Tunnel Barrier in the Strong Coupling Regime. *Proc. Natl. Acad. Sci. U. S. A.* **2012**, *109*, 11498–11503.

(5)    Yan, H.; Bergren, A. J.; McCreery, R.; Della Rocca, M. L.; Martin, P.; Lafarge, P.; Lacroix, J.-C. C.; Rocca, M. L. Della; Martin, P.; Lafarge, P.; *et al.* Activationless Charge Transport across 4.5 to 20nm in Molecular Electronic Junctions. *Proc. Natl. Acad. Sci. U. S. A.* **2013**, *110*, 5326–5330.

(6)    Mujica, V.; Ratner, M. A.; Nitzan, A. Molecular Rectification: Why Is It so Rare? *Chem. Phys.* **2002**, *281*, 147–150.

(7)    Van Dyck, C.; Ratner, M. A. Molecular Rectifiers: A New Design Based on Asymmetric Anchoring Moieties. *Nano Lett.* **2015**, *15*, 1577–1584.

(8)    Aviram, A.; Ratner, M. A. Molecular Rectifiers. *Chem. Phys. Lett.* **1974**, *29*, 277–283.

(9)    Metzger, R. M. Unimolecular Electronics. *Chem. Rev.* **2015**, *115*, 5056–5115.

(10)  Chen, X.; Roemer, M.; Yuan, L.; Du, W.; Thompson, D.; del Barco, E.; Nijhuis, C. A. Molecular Diodes with Rectification Ratios Exceeding 105 Driven by Electrostatic Interactions. *Nat. Nanotechnol.* **2017**, *12*, 797–803.

(11)  Yuan, L.; Nerngchamnong, N.; Cao, L.; Hamoudi, H.; del Barco, E.; Roemer, M.; Sriramula, R. K.; Thompson, D.; Nijhuis, C. A. Controlling the Direction of Rectification in a Molecular Diode. *Nat. Commun.* **2015**, *6*, 6324.

(12)  Bunch, J. S.; Verbridge, S. S.; Alden, J. S.; van der Zande, A. M.; Parpia, J. M.; Craighead, H. G.; McEuen, P. L. Impermeable Atomic Membranes from Graphene Sheets. *Nano Lett.* **2008**, *8*, 2458–2462.

(13)  Lee, C.-H.; Schiros, T.; Santos, E. J. G.; Kim, B.; Yager, K. G.; Kang, S. J.; Lee, S.; Yu, J.; Watanabe, K.; Taniguchi, T.; *et al.* Epitaxial Growth of Molecular Crystals on van Der Waals Substrates for High-Performance Organic Electronics. *Adv. Mater.* **2014**, *26*, 2812–2817.

(14)  Leitherer, S.; Coto, P. B.; Ullmann, K.; Weber, H. B.; Thoss, M. Charge Transport in C $_{60}$ -Based Single-Molecule Junctions with Graphene Electrodes. *Nanoscale* **2017**, *9*, 7217–7226.

(15)  Jang, Y.; Jeong, H.; Kim, D.; Hwang, W.-T.; Kim, J.-W.; Jeong, I.; Song, H.; Yoon, J.; Yi, G.-C.; Jeong, H.; *et al.* Electrical Characterization of Benzenedithiolate Molecular Electronic Devices with Graphene Electrodes on Rigid and Flexible Substrates. *Nanotechnology* **2016**, *27*, 145301.

(16)  Jang, Y.; Kwon, S. J.; Shin, J.; Jeong, H.; Hwang, W. T.; Kim, J.; Koo, J.; Ko, T. Y.; Ryu, S.; Wang, G.; *et al.* Interface-Engineered Charge-Transport Properties in Benzenedithiol Molecular Electronic Junctions via Chemically P-Doped Graphene Electrodes. *ACS Appl. Mater. Interfaces* **2017**, *9*, 42043–42049.

(17) Island, J. O.; Holovchenko,  a; Koole, M.; Alkemade, P. F. a; Menelaou, M.; Aliaga-Alcalde, N.; Burzurí, E.; van der Zant, H. S. J. Fabrication of Hybrid Molecular Devices Using Multi-Layer Graphene Break Junctions. *J. Phys. Condens. Matter* **2014**, *26*, 474205.

(18) Prins, F.; Barreiro, A.; Ruitenberg, J. W.; Seldenthuis, J. S.; Aliaga-Alcalde, N.; Vandersypen, L. M. K.; van der Zant, H. S. J. Room-Temperature Gating of Molecular Junctions Using Few-Layer Graphene Nanogap Electrodes. *Nano Lett.* **2011**, *11*, 4607–4611.

(19) Burzurí, E.; Island, J. O.; Díaz-Torres, R.; Fursina, A.; González-Campo, A.; Roubeau, O.; Teat, S. J.; Aliaga-Alcalde, N.; Ruiz, E.; Van Der Zant, H. S. J. Sequential Electron Transport and Vibrational Excitations in an Organic Molecule Coupled to Few-Layer Graphene Electrodes. *ACS Nano* **2016**, *10*, 2521–2527.

(20) Zhang, Q.; Liu, L.; Tao, S.; Wang, C.; Zhao, C.; González, C.; Dappe, Y. J.; Nichols, R. J.; Yang, L. Graphene as a Promising Electrode for Low-Current Attenuation in Nonsymmetric Molecular Junctions. *Nano Lett.* **2016**, *16*, 6534–6540.

(21) Rudnev, A. V.; Kaliginedi, V.; Droghetti, A.; Ozawa, H.; Kuzume, A.; Haga, M. aki; Broekmann, P.; Rungger, I. Stable Anchoring Chemistry for Room Temperature Charge Transport through Graphite-Molecule Contacts. *Sci. Adv.* **2017**, *3*.

(22) Kim, T.; Liu, Z.-F.; Lee, C.; Neaton, J. B.; Venkataraman, L. Charge Transport and Rectification in Molecular Junctions Formed with Carbon-Based Electrodes. *Proc. Natl. Acad. Sci.* **2014**, *111*, 10928–10932.

(23) Martin, P.; Della Rocca, M. L.; Anthore, A.; Lafarge, P.; Lacroix, J.-C. Organic Electrodes Based on Grafted Oligothiophene Units in Ultrathin, Large-Area Molecular Junctions. *J. Am. Chem. Soc.* **2012**, *134*, 154–157.

(24) Liang, X.; Sperling, B. A.; Calizo, I.; Cheng, G.; Hacker, C. A.; Zhang, Q.; Obeng, Y.; Yan, K.; Peng, H.; Li, Q.; *et al.* Toward Clean and Crackless Transfer of Graphene. *ACS Nano* **2011**, *5*, 9144–9153.

(25) Ferrari, A. C.; Meyer, J. C.; Scardaci, V.; Casiraghi, C.; Lazzeri, M.; Mauri, F.; Piscanec, S.; Jiang, D.; Novoselov, K. S.; Roth, S.; *et al.* Raman Spectrum of Graphene and Graphene Layers. *Phys. Rev. Lett.* **2006**, *97*, 187401.

(26) Mesnage, A.; Lefèvre, X.; Jégou, P.; Deniau, G.; Palacin, S. Spontaneous Grafting of Diazonium Salts: Chemical Mechanism on Metallic Surfaces. *Langmuir* **2012**, *28*, 11767–11778.

(27) Fave, C.; Leroux, Y.; Trippé, G.; Randriamahazaka, H.; Noel, V.; Lacroix, J.-C.; Trippé, G.; Randriamahazaka, H.; Noel, V.; Lacroix, J.-C. Tunable Electrochemical Switches Based on Ultrathin Organic Films. *J. Am. Chem. Soc.* **2007**, *129*, 1890–1891.

(28) Fave, C.; Noel, V.; Ghilane, J.; Trippé-Allard, G.; Randriamahazaka, H.; Lacroix, J.-C. Electrochemical Switches Based on Ultrathin Organic Films: From Diode-like Behavior to Charge Transfer Transparency. *J. Phys. Chem. C* **2008**, *112*, 18638–18643.

(29) Greenwood, J.; Phan, T. H.; Fujita, Y.; Li, Z.; Ivasenko, O.; Uji-i, H.; Mertens, S. F. L.; Feyter, S. De. Covalent Modification of Graphene and Graphite Using Diazonium Chemistry : Tunable Grafting and Nanomanipulation. *ACS Nano* **2015**, *9*, 5520–5535.

(30) Huang, S.; Ling, X.; Liang, L.; Song, Y.; Fang, W.; Zhang, J.; Kong, J.; Meunier, V.; Dresselhaus, M. S. Molecular Selectivity of Graphene-Enhanced Raman Scattering. *Nano Lett.* **2015**, *15*, 2892–2901.

(31)  Kupka, T.; Wrzalik, R.; Pasterna, G.; Pasterny, K. Theoretical DFT and Experimental Raman and NMR Studies on Thiophene, 3-Methylthiophene and Selenophene. *J. Mol. Struct.* **2002**, *616*, 17–32.

(32)  Venugopal, A.; Colombo, L.; Vogel, E. M. Contact Resistance in Few and Multilayer Graphene Devices. *Appl. Phys. Lett.* **2010**, *96*, 13512.

(33)  Ishii, H.; Sugiyama, K.; Ito, E.; Seki, K. Energy Level Alignment and Interfacial Electronic Structures at Organic/Metal and Organic/Organic Interfaces. *Adv. Mater.* **1999**, *11*, 605–625.

(34)  Baldo, M.; Forrest, S. Interface-Limited Injection in Amorphous Organic Semiconductors. *Phys. Rev. B* **2001**, *64*, 85201.

(35)  Vasquez, H.; Oszwaldowski, R.; Pou, P.; Ortega, J.; Flores, F.; Kahn, A. Dipole Formation at metal/PTCDA Interfaces: Role of the Charge Neutrality Level. *Europhys. Lett.* **2004**, *65*, 802.

(36)  Droghetti, A.; Thielen, P.; Rungger, I.; Haag, N.; Großmann, N.; Stöckl, J.; Stadtmüller, B.; Aeschlimann, M.; Sanvito, S.; Cinchetti, M. Dynamic Spin Filtering at the Co/Alq3 Interface Mediated by Weakly Coupled Second Layer Molecules. *Nat. Commun.* **2016**, *7*, 12668.

(37)  Nguyen, Q. van; Martin, P.; Frath, D.; Della Rocca, M. L.; Lafolet, F.; Barraud, C.; Lafarge, P.; Mukundan, V.; James, D.; McCreery, R. L.; *et al.* Control of Rectification in Molecular Junctions: Contact Effects and Molecular Signature. *J. Am. Chem. Soc.* **2017**, *139*, 11913–11922.

(38) Braun, S.; Salaneck, W. R.; Fahlman, M. Energy-Level Alignment at Organic/Metal and Organic/Organic Interfaces. *Adv. Mater.* **2009**, *21*, 1450–1472.

(39) Kahn, A.; Koch, N.; Gao, W. Electronic Structure and Electrical Properties of Interfaces between Metals and Pi-Conjugated Molecular Films. *J. Polym. Sci. Part B Polym. Phys.* **2003**, *41*, 2529–2548.

(40) Alloway, D.; Hofmann, M.; Smith, D. L.; Gruhn, N. E.; Graham, A.; Colorado, R.; Wysocki, V. H.; Lee, T. R.; Lee, P.; Armstrong, N. R. Interface Dipoles Arising from Self-Assembled Monolayers on Gold: UV−Photoemission Studies of Alkanethiols and Partially Fluorinated Alkanethiols. *J. Phys. Chem. N* **2003**, *107*, 11690–11699.